\documentclass[12pt]{article}
\usepackage{graphicx}
\usepackage{epsfig}
\usepackage{amssymb,amsmath}
\usepackage{pstricks}

\input{epsf.sty}

\newcommand{\be}{\begin{equation}}
\newcommand{\ee}{\end{equation}}
\newcommand{\ba}{\begin{eqnarray}}
\newcommand{\ea}{\end{eqnarray}}
\newcommand{\bi}{\begin{itemize}}
\newcommand{\ei}{\end{itemize}}

\newcommand{\nn}{\nonumber}

\newcommand{\RR}{{\rm I\kern -.2em  R}} 
\newcommand{\eq}{Eq.~}
\newcommand{\eqs}{Eqs.~}
\newcommand{\fig}{Fig.~}

\def\lsi{\raise0.3ex\hbox{$<$\kern-0.75em\raise-1.1ex\hbox{$\sim$}}}
\def\gsi{\raise0.3ex\hbox{$>$\kern-0.75em\raise-1.1ex\hbox{$\sim$}}}
\newcommand{\lsim}{\mathop{\lsi}}

\def\none               {\multicolumn{2}{c|}{---}}

\def\slash#1{#1\!\!\!\!/\!\,\,}
\def\Dslash{\slash D}

\begin{document}
 
\begin{titlepage}
\begin{flushright}
CERN-PH-TH/2008-152\\
MS-TP-08-15
\end{flushright}
\begin{centering}
\vfill
                   
{\bf\Large The chiral critical point of $N_f=3$ QCD \\ at finite density
           to the order $(\mu/T)^4$}

\vspace{0.8cm}

Philippe de Forcrand$^{1,2}$ and Owe Philipsen$^3$

\vspace{0.3cm}
{\em $^{\rm 1}$
Institut f\"ur Theoretische Physik,
ETH Z\"urich,
CH-8093 Z\"urich,
Switzerland\\}
{\em $^{\rm 2}$
CERN, Physics Department, TH Unit, CH-1211 Geneva 23,
Switzerland\\}
{\em $^{\rm 3}$
Institut f\"ur Theoretische Physik, Westf\"alische Wilhelms-Universit\"at M\"unster, Germany}

\vspace*{0.7cm}
 
\begin{abstract}
QCD with three degenerate quark flavours at zero baryon density exhibits a first order thermal phase 
transition for small quark masses, which changes to a smooth crossover for
some critical quark mass $m^c_0$, i.e.~the {\em chiral} critical point. It is generally believed
that as an (even) function of quark chemical potential, the critical point $m_c(\mu)$ moves
to larger quark masses, constituting the critical endpoint of a first order phase
transition in theories with $m\geq m^c_0$. 
To test this, we consider a Taylor expansion of $m_c(\mu)$
around $\mu=0$ and determine the first two coefficients from lattice simulations
with staggered fermions on $N_t=4$ lattices. We employ two
different techniques: a) calculating the coefficients directly from a $\mu=0$ ensemble 
using a novel finite difference method, and 
b) fitting them to simulation data obtained for imaginary chemical potentials.
The $\mu^2$ and $\mu^4$ coefficients are found to be negative by both methods, with consistent
absolute values. Combining both methods gives evidence that also the 
$\mu^6$ coefficient is negative. Hence, on coarse $N_t=4$ lattices a three-flavour theory  with 
$m> m^c_0$ does not possess a {\em chiral} critical endpoint for quark chemical potentials $\mu\lsim T$.
Simulations on finer lattices are required for reliable continuum physics. Possible implications
for the QCD phase diagram are discussed.

\end{abstract}
\end{centering}
 
\noindent
\vfill
\noindent
 

\vfill

\end{titlepage}
 

\section{Introduction}

The understanding of the QCD phase diagram as a function of temperature and baryon 
density plays an important role in an increasing
number of nuclear and particle physics programmes, with possible applications to
cosmology and astro-particle physics. Of particular interest in the context of 
heavy ion collision experiments is the location and nature of the quark-hadron 
transition. The generally expected picture -- with an analytical
crossover at $\mu=0$ turning into a first order transition
beyond some critical chemical potential $\mu_E$ -- is based on
combining lattice results for $\mu=0$ with model calculations at $\mu\neq 0$
and connecting various limiting cases by universality and
continuity arguments \cite{derivs}.  
In particular it is generally believed that, in the enlarged parameter space $\{m_{u,d},m_s,T,\mu\}$,
the critical endpoint of QCD with physical quark masses is
continuously connected to the chiral critical line at $\mu_E=0$, which delimits the quark mass region 
with a first order chiral phase transition at zero density. 

\begin{figure}[t]
\includegraphics[width=0.28\textwidth]{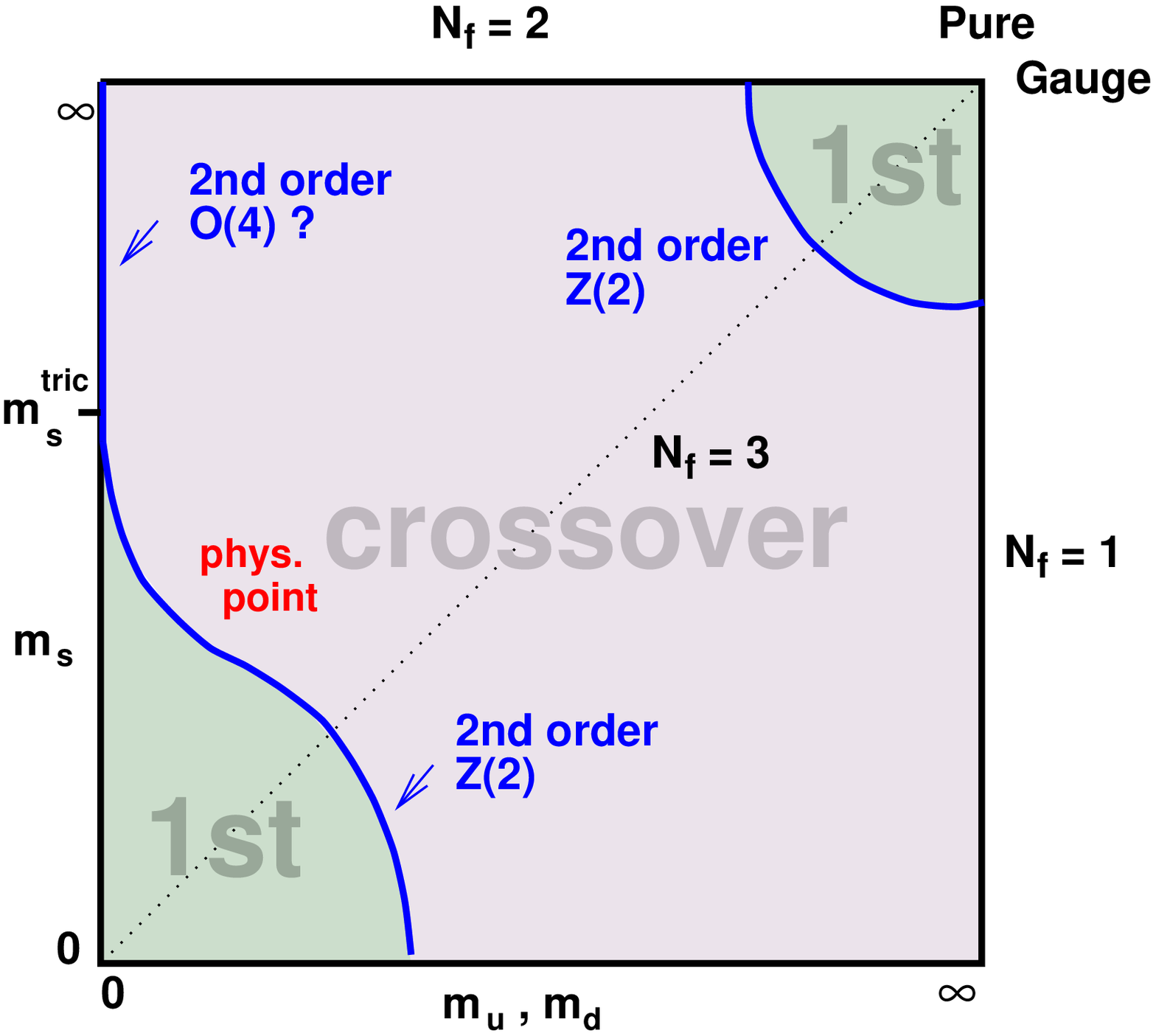}
\includegraphics[width=0.40\textwidth]{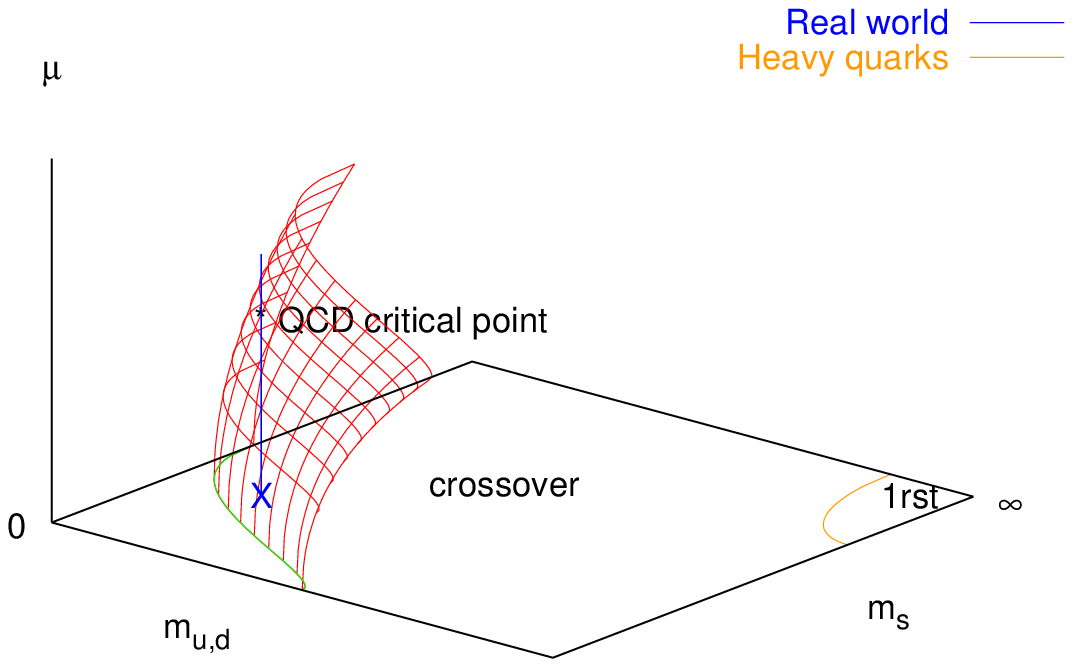}
\put(-125.5,64.5){\tiny $\bullet$}
\includegraphics[width=0.40\textwidth]{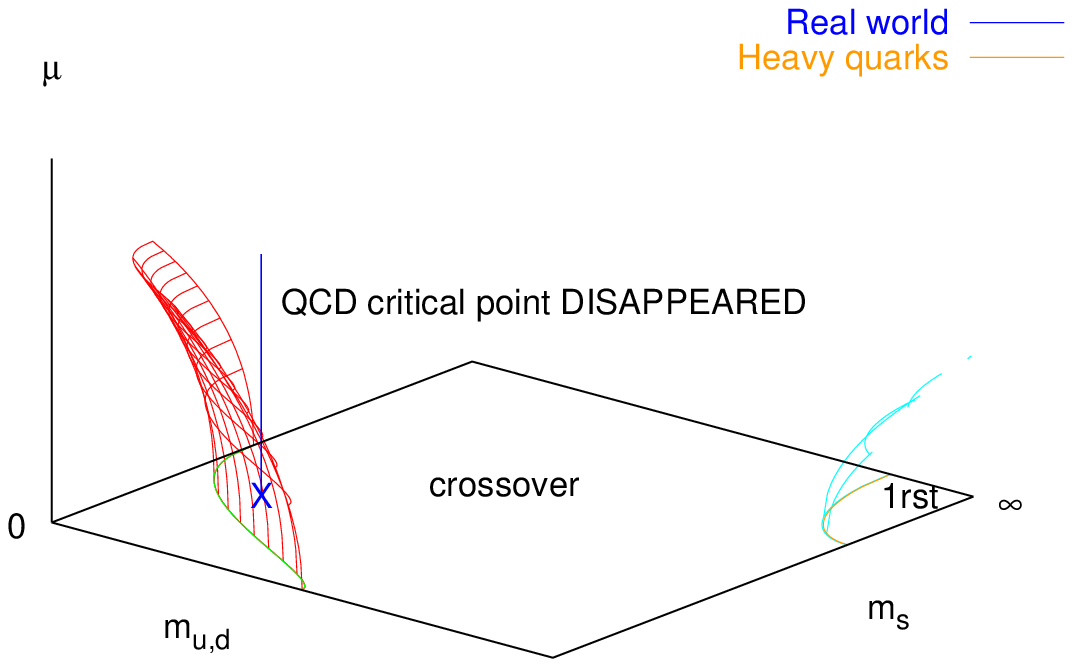}
\caption[]{\label{schem} Left: Schematic phase transition behaviour of $N_f=2+1$
QCD for different choices of quark masses $(m_{u,d},m_s)$ at
$\mu=0$. Middle/Right: Critical surface swept by the chiral 
critical line as $\mu$ is turned on. Depending on the curvature, a QCD chiral critical
point is present or absent \cite{oprev}. 
For heavy quarks the curvature has been determined \cite{kim1} and the first order 
region shrinks with $\mu$.

}
\end{figure}

The schematic situation at $\mu=0$ is depicted in \fig\ref{schem} (left).
In the limits of zero and infinite quark masses (lower left and upper
right corners), order parameters corresponding to the breaking of a
global symmetry can be defined, and one numerically finds first-order phase
transitions at small and large quark masses at some finite
temperatures $T_c(m)$. On the other hand, one observes an analytic crossover at
intermediate quark masses, with second order boundary lines separating these
regions. Both lines have been shown to belong to the $Z(2)$ universality class
of the $3d$ Ising model \cite{kls,fp2,kim1}. The chiral critical line
has been mapped out on $N_t=4$ lattices, and the physical point is confirmed
to be on the crossover side of this line \cite{fp3,nature}.
The continuity arguments in \cite{derivs} 
suggest that this line should shift with $\mu$ towards larger quark masses, thus increasing the
region of a chiral first order transition. 
At some value $\mu_E$ it passes through the physical point,  
whence it represents the endpoint of the phase diagram for physical QCD, \fig\ref{schem} (middle).

This scenario has recently been called into question by numerical simulations at imaginary chemical
potential $\mu=i\mu_i, \mu_i\in \Bbb{R}$ \cite{fp3}. 
There it was found that, as a function of $\mu_i$,
the chiral critical line moves towards larger quark masses. 
In the degenerate $N_f=3$ theory and in the neighbourhood of the critical point 
$m_{u,d}^c=m_s^c\equiv m^c_0$, the dependence 
of the critical temperature and quark mass on the chemical potential can be described by
a Taylor expansion, with only even powers of $\mu$ due to CP-symmetry,
\ba
\label{be}
\frac{T_c(m,\mu)}{T_c(m^c_0,0)}&=&1+\sum_{k,l=1} \alpha_{kl}
\left(\frac{m-m^c_0}{\pi T_c}\right)^k
\label{mc1}
\left(\frac{\mu}{\pi T_c}\right)^{2l},\\
\frac{m_c(\mu)}{m_c(0)}&=&1+\sum_{k=1} c_{k} \left(\frac{\mu}{\pi T_c}\right)^{2k}.
\label{mc}
\ea
By performing simulations
at several values of $\mu_i$ and quark masses, the leading coefficients were determined, and
in particular $c_1$ was found to be negative. This implies a 
scenario as in \fig\ref{schem} (right), where the chiral critical line does not 
cross the physical point
for small (real) chemical potentials. If this behaviour continues to larger values of $\mu$, there 
would be no non-analytic chiral phase transition or end point at all in physical QCD.

Clearly, it is now important to assess the reliability of this result. The main systematic 
uncertainties associated with the calculation in \cite{fp3} are the truncation of 
the Taylor expansion \eq(\ref{mc}) and the coarse lattice with $N_t=4, a\sim 0.3$ fm.
In this letter we perform a systematic investigation of the former. 
Firstly, instead of extracting the leading term in \eq(\ref{mc}) 
from fits to a truncated Taylor series,
we instead calculate the derivative with respect to $\mu^2$ directly. Our calculation
confirms the negative sign of the leading term on two different volumes and with drastically 
reduced errors.
Second, we repeat our fits to Taylor series with 
vastly increased statistics and simulations at additional values of chemical potentials. This 
allows us to also constrain the ${\cal O}(\mu^4)$ and the sign of the ${\cal O}(\mu^6)$ terms, which 
we both find to be negative as well. Moreover, our leading
and next-to-leading order coefficients agree within statistical 
errors between the two methods. Our findings leave little doubt that on coarse $N_t=4$ lattices with
staggered fermions the scenario \fig\ref{schem} (right) is realised.

\section{Extracting the $\mu$ dependence of the critical point}

Our strategy to determine the $\mu$ dependence of the critical quark masses has been
described in detail in \cite{fp3}. In order to keep this paper self-contained, we repeat
the essential formulae.
On the lattice, the critical parameters \eqs(\ref{mc1},\ref{mc}) get replaced by critical 
couplings, which in turn have a Taylor expansion around the zero density critical quark mass,
\ba
\label{bc}
\beta_c(am,a\mu) & = & \beta_c(am^c_0,0)+ \sum_{k,l=1} c_{kl}\, (am-am^c_0)^k \, (a\mu)^{2l}, \\
\label{mclat}
am^c(a\mu)& = & am^c_0 + \sum_{k=1} c'_k \, (a\mu)^{2k}\;.
\ea
As in our previous investigations, our observable for criticality is the Binder cumulant constructed from
the fluctuations of the chiral condensate,
\be
B_4 \equiv \frac{\langle (\delta \bar{\psi}\psi)^4 \rangle}{\langle (\delta \bar{\psi}\psi)^2 \rangle^2},
\quad \delta \bar{\psi}\psi = \bar{\psi}\psi - \langle \bar{\psi}\psi \rangle.
\ee
In the infinite volume limit the Binder cumulant behaves discontinuously, 
assuming the values 1 in a first order regime,
3 in a crossover regime and 1.604 characteristic of the $3d$ Ising universality 
at a second order chiral critical point. 
In a finite volume the discontinuities are smeared out and flattened,  
so that $B_4$ passes continuously through the critical value, i.e.~it is an analytic function of
$(\beta,am,a\mu)$. For a given pair $(am,a\mu)$, the gauge coupling needs to be 
fixed to its pseudo-critical value $\beta_c(am,a\mu)$ by requiring,
for instance, that the susceptibility $\langle (\delta X)^2 \rangle$ be maximum. We use the equivalent
prescription of requiring a vanishing third moment, $M_3 \equiv \langle (\delta X)^3 \rangle = 0$. In the neighbourhood of the chiral critical 
point  the Binder cumulant can then be expanded as
\begin{equation}
B_4(am,a\mu)=1.604+\sum_{k,l=1}b_{kl}\, (am-am^c_0)^k(a\mu)^{2l}\;,
\label{bseries}
\end{equation}
with volume dependent coefficients $b_{kl}(L)$.

For large volumes the approach to the thermodynamic
limit is governed by universality.
Near a critical point the correlation length diverges
as $\xi\sim r^{-\nu}$, where $r$ is the distance to the critical point
in the plane of temperature and magnetic field-like variables, and $\nu\approx 0.63$
for the $3d$ Ising universality class. 
Since $\beta$ is tuned to $\beta_c$ always, we have $r\propto |am-am^c_0|$ for $\mu=0$ 
and $r\propto |a\mu|^2$ for $am=am^c_0$ .
$B_4$ is a function of the dimensionless ratio $L/\xi$, or equivalently
$(L/\xi)^{1/\nu}$. Hence, for the Taylor coefficients  one expects, for large volumes,
the universal scaling behaviour 
\be \label{scale}
b_{kl}(L)=f_{kl} L^{(k+l)/\nu}\;,
\ee
where the $f_{kl}$ are independent of the volume.

It is now straightforward to relate the desired coefficients in \eq(\ref{mclat}) 
to those appearing in the Binder cumulant \eq(\ref{bseries}). The function $am^c(a\mu)$ is defined
implicitly by the condition that the Binder cumulant stays critical, $B_4(am,a\mu)=1.604$.
Taking total derivatives with respect to $\mu^2$ one obtains
\ba 
c'_1&=&\frac{d\,am^c}{d(a\mu)^2}=-\frac{\partial B_4}{\partial (a\mu)^2}
\left(\frac{\partial B_4}{\partial am}\right)^{-1}=-\frac{b_{01}}{b_{10}} \quad , \nn\\
c'_2&=&
\frac{1}{2!}\,\frac{d^2\,am^c}{d[(a\mu)^2]^2}
=-\frac{1}{b_{10}}(b_{02}+b_{11}c'_1+b_{20}{c'_1}^2) \quad , \nn\\
c'_3&=&\frac{1}{3!}\,\frac{d^3\,am^c}{d[(a\mu)^2]^3}
=-\frac{1}{b_{10}}(b_{03}+b_{11}c'_2+b_{12}c'_1+b_{21}{c'_1}^2+2b_{20}c'_1c'_2+b_{30}{c'_1}^3) \quad .
\label{der2}
\ea
Once these coefficients are known, they need to be converted to the continuum
coefficients \eq(\ref{mc}). This is non-trivial:  for fixed temporal lattice extent $N_t$ the lattice 
spacings entering the dimensionless $am^c(\mu)$ and $am^c(0)$ are different, 
since the pseudo-critical temperature $T_c(m_c(\mu),\mu)=1/(N_t a(\mu))$ depends on $\mu$. 
The relation between the continuum and lattice coefficients is thus
\ba
c_1&=&\frac{\pi^2 }{N_t^2}\, \frac{c_1'}{am_0^c}
+\frac{1}{T_c(m^c_0,0)}\frac{dT_c(m^c(\mu),\mu)}{d(\mu/\pi T)^2},\nn\\
c_2&=&
\frac{\pi^4}{N_t^4}\,\frac{c'_2}{am^c_0}
-\frac{\pi^2}{N_t^2}\,\frac{c'_1}{am^c_0}\,
\frac{1}{T_c(m^c_0,0)}\frac{dT_c(m^c(\mu),\mu)}{d(\mu/\pi T)^2}\nn\\
&&
+\frac{1}{2T_c(m^c_0,0)}\,\frac{d^2T_c(m^c(\mu),\mu)}{d[(\mu/\pi T)^2]^2}\;.
\label{conv2}
\ea
The pseudo-critical temperature $T_c(m,\mu)$ can be obtained from $\beta_c(am,a\mu)$ by means
of the two-loop beta-function. Note that for our coarse lattices $N_t=4$ this procedure is
certainly not quantitatively valid, and some non-perturbative beta-function should be used.
However, we shall see that for the conclusions to be drawn later the two-loop beta-function is
actually sufficient.

With these equations, we are ready for numerical evaluations. Our task is to extract the 
coefficients $b_{kl}$ of the Binder cumulant expansion, \eq(\ref{bseries}), from 
numerical simulations, and then convert them to the continuum via 
\eqs(\ref{der2},\ref{conv2}). In the following sections we  
present results for two independent ways of measuring the $b_{kl}$.

\section{Calculating derivatives \label{delta}}

When attempting to extract Taylor coefficients by fits of truncated series to data containing
the full $\mu$-dependence at imaginary $\mu$, one has to worry about systematic errors. 
Fitting to a finite polynomial necessarily distributes the actual $\mu$-dependence across
the available terms, and the result is only reliable to the extent that the higher terms are
numerically negligible. In particular, leading order fits run a risk of  ``averaging'' a more complicated functional dependence into a single coefficient, especially when the function is very flat. 
These dangers have recently been demonstrated
for the pseudo-critical coupling in SU(2), where calculations at real $\mu$ 
are possible and can be compared to results obtained via analytic continuation from 
imaginary $\mu$ \cite{bari}.  

To guard against this, and because the $\mu$-dependence of $B_4$ is 
extremely weak \cite{fp3}, we now wish to calculate the derivative $\partial B_4/\partial (a\mu)^2$ directly, 
without recourse to fitting. (For $\partial B_4/\partial (am)$ this is not necessary, the $m$-dependence of $B_4$
is pronounced enough to give reliable fits \cite{fp2,fp3}). 
In principle, this can be done by expressing the 
$\mu-$derivative of $B_4$ analytically through traces of non-local operators,
whose expectation values are to be evaluated at $\mu=0$. 
This technique has been used to calculate the Taylor coefficients of the pseudo-critical 
temperature, the pressure and various susceptibilities in \cite{biswa,gg}. 
However, we do not follow this approach here.
First, the numerical effort is non-trivial. For instance, the trace of the $5^{th}$
inverse power of the Dirac operator must be evaluated to obtain the derivative 
of $(\delta\bar\psi \psi)^4$ with respect to the quark mass. 
Moreover, delicate cancellations will take place among the rapidly growing number of terms, leading 
to difficulties with optimising the number of noise vectors to be used as accurate
stochastic estimators of the various traces.

Instead, we use a  new method to estimate the derivative, which turns out to be simpler to implement 
and vastly more efficient.
We measure the change $\Delta B_4$ under a
small variation $\Delta(a\mu)^2$,
thus estimating the ratio of finite differences
which, for sufficiently small variations, approaches the desired derivative,  
\be
\lim_{\Delta(a\mu^2)\rightarrow 0}\frac{\Delta B_4}{\Delta (a\mu)^2}=
\left.\frac{\partial B_4}{\partial (a\mu)^2}\right|_{\mu=0}.
\ee 
Because the required shift in the couplings is very small,
it is adequate and safe to use the original Monte Carlo ensemble for
$am^c_0, \mu=0$ and reweight the results by the standard 
Ferrenberg-Swendsen method~\cite{reweight}. Moreover, 
reweighting to imaginary $\mu$
the reweighting factors remain real positive and close to 1.
With reweighting, the fluctuations in the original and the
reweighted ensembles are strongly correlated. For the calculation of variations
this is beneficial. The correlated fluctuations drop out
of our observable, which now is 
the {\em change} in $B_4$, rather than $B_4$ itself.
As a final simplification, we do not calculate the reweighting
factors 
\be
\rho(\mu_1,\mu_2)=\frac{\det^{N_f/4} \Dslash(U,\mu_2)}
{\det^{N_f/4}\Dslash(U,\mu_1)} 
\ee
exactly, but estimate them through Gaussian noise vectors $\eta$ in the standard way,
\be
\rho(\mu_1,\mu_2)= \left\langle \exp\left(-|\Dslash^{-N_f/8}(\mu_2) 
\Dslash^{+N_f/8}(\mu_1) \eta|^2 + |\eta|^2 \right) \right\rangle_\eta .
\ee
As a further check of the method on a more accurately calculable observable, 
we also consider the pseudo-critical coupling $\beta_c$
defined by the peak of the plaquette susceptibility,   
and calculate $\partial\beta_c/\partial (a\mu)^2$ along the way. 
Further details and a thorough test of this method in the Potts model are described
in \cite{LAT07},
where the results were found to fully agree with the other method of 
computing derivatives. Since that test the technique has also been successfully 
applied in a similar calculation at finite isospin chemical potential \cite{ks}.

\begin{figure}[t]
\hspace*{-1cm}
\includegraphics[width=0.55\textwidth]{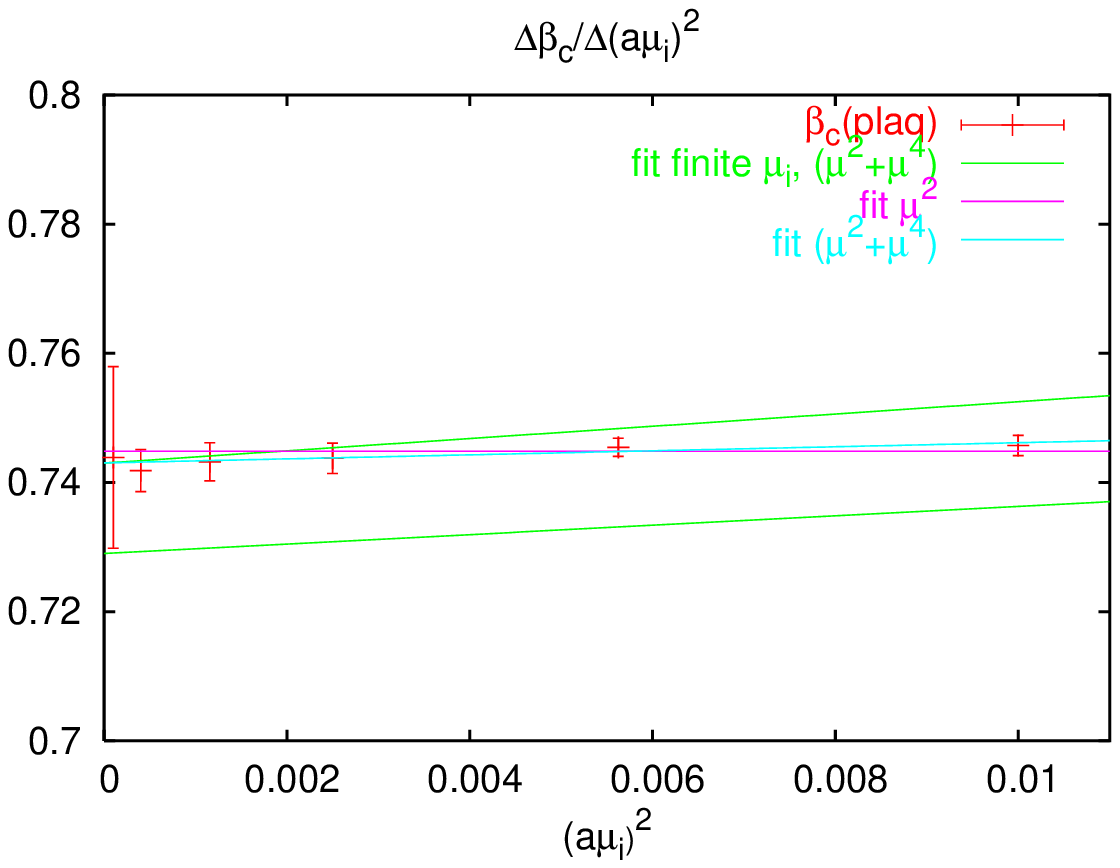}
\includegraphics[width=0.55\textwidth]{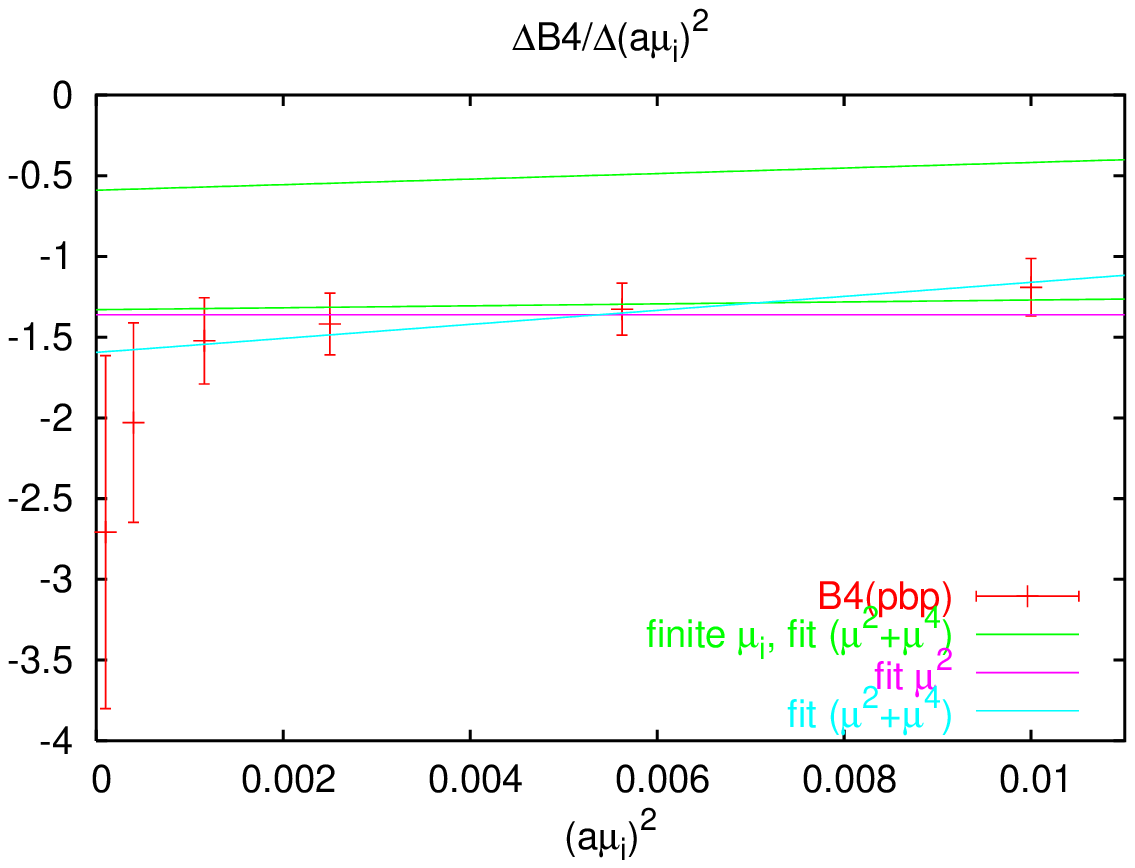}
\caption{\label{der} Finite difference quotients on $L=8$ extrapolated to zero as a constant, 
or constant plus $(a\mu_i)^2$
correction. The resulting coefficients are listed in Table \ref{tabder}. The error band corresponds 
to the highlighted best fits to data at imaginary $\mu$, Table \ref{B4table}. 
}
\end{figure}

Our calculations for $\partial \beta_c/\partial (a\mu)^2, \partial B_4/\partial (a\mu)^2$ were done on
$8^3\times 4$ and $12^3\times 4$ lattices, for which we have accumulated about
5 million and 0.5 million trajectories, respectively
\footnote{The $8^3\times 4$ simulations were performed in about two months on the EGEE computing grid, 
thanks to the support of the CERN IT/GS group. Jobs at different parameter sets were distributed 
on the grid infrastructure where typically $\sim 1000$ CPUs could be used at the same time 
for this activity.}. 
The bare quark mass was kept fixed at $am=0.0265$, which is close to the critical value $am^c_0$
and corresponds to $m_\pi L \sim 3.4, 5$, respectively.
(A small offset from $am_0^c$ will affect the $\mu$-derivative only via the lowest order mixing term,
whose coefficient is compatible with zero, as we shall see in the next section.)
The results of this calculation are shown in \fig\ref{der}.
The individual data points correspond to the difference quotient reweighted to the given
value of $\mu_i$. The error bars within one plot are thus strongly correlated.
The straight lines represent linear extrapolations to zero $\mu_i$, 
and the intercept is the final estimate for the respective derivatives. 
The extrapolations with non-zero slope 
show the influence of the next-to-leading term. 
All estimates are collected in Table \ref{tabder}. 

\begin{table}[t]
\begin{center}
\hspace*{-1cm}
\begin{tabular}{|*{3}{r@{.}l|}|*{5}{r@{.}l|}}
\hline
\multicolumn{2}{|c|}{$c_{01}$} &
\multicolumn{2}{c|}{$c_{02}$} &
\multicolumn{2}{c||}{$\chi^2/{\rm dof}$} &
\multicolumn{2}{c|}{$b_{01}$} &
\multicolumn{2}{c|}{$b_{02}$} &
\multicolumn{2}{c|}{$f_{01}$} &
\multicolumn{2}{c|}{$f_{02}$} &
\multicolumn{2}{c|}{$\chi^2/{\rm dof}$} \\
\hline 
  0&7448(6)  & \none      & 0&38 &  1&36(8)  & \none        &   0&050(3) & \none & 0&82  \\
   0&7430(6)  & 0&31(9)    & 0&12 &  1&59(12) & 43&4(18.8) & 0&059(4) & 0&059(26)&  0&33 \\ 
   \hline
 0&737(3)   & \none      & 1&10 &  2&63(40) & \none        & 0&051(8) & \none & 0&50 \\  0&7430(6)  & 0&14(1.10) & 1&37 &  3&35(53) & 259&0(148.0) & 0&065(10) & 0&097(56) & 0&44 \\
\hline
\end{tabular}
\end{center}
\caption[]{\label{tabder} Taylor coefficients of 
$\beta_c(am,a\mu)$ and $B_4(am,a\mu)$, \eqs (\ref{bc},\ref{bseries},\ref{scale}), 
from the extrapolated finite differences, as shown in \fig\ref{der}. The upper rows are for $L=8$,
the lower rows for $L=12$.
}
\end{table}                                         

\begin{figure}[t]
\hspace*{-1cm}
\includegraphics[width=0.55\textwidth]{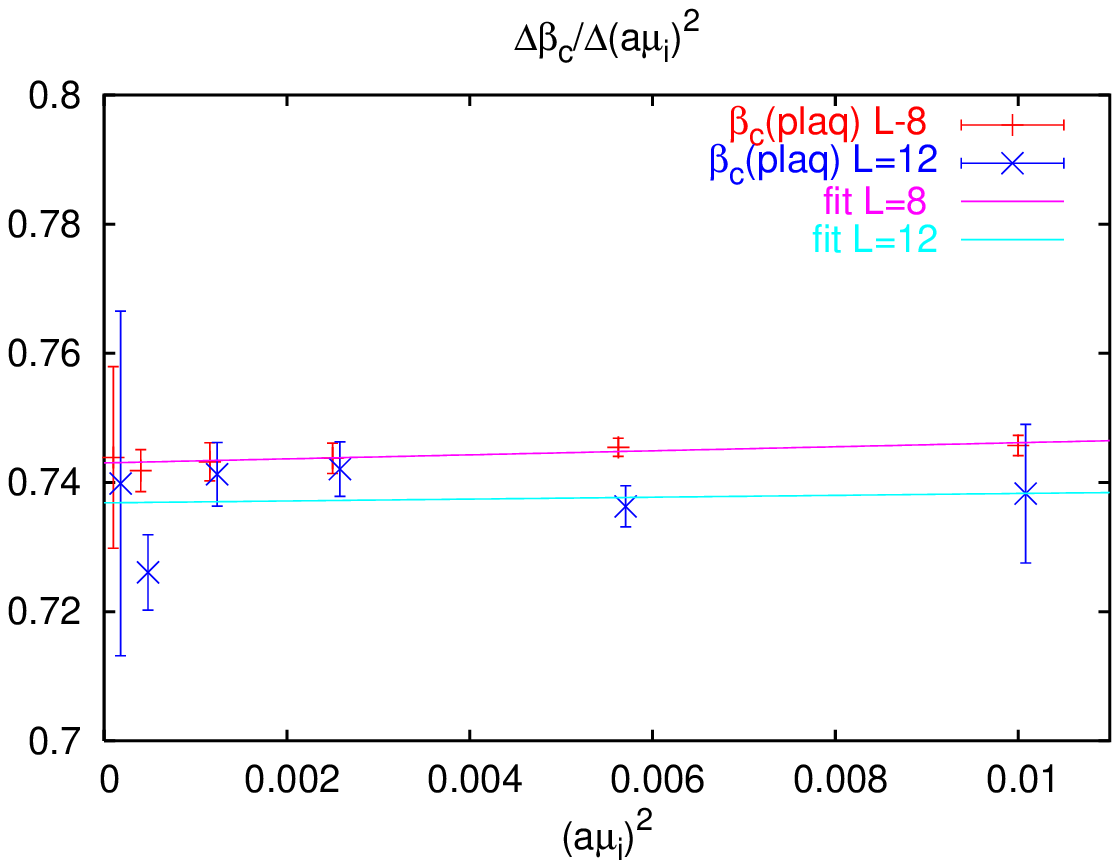}
\includegraphics[width=0.55\textwidth]{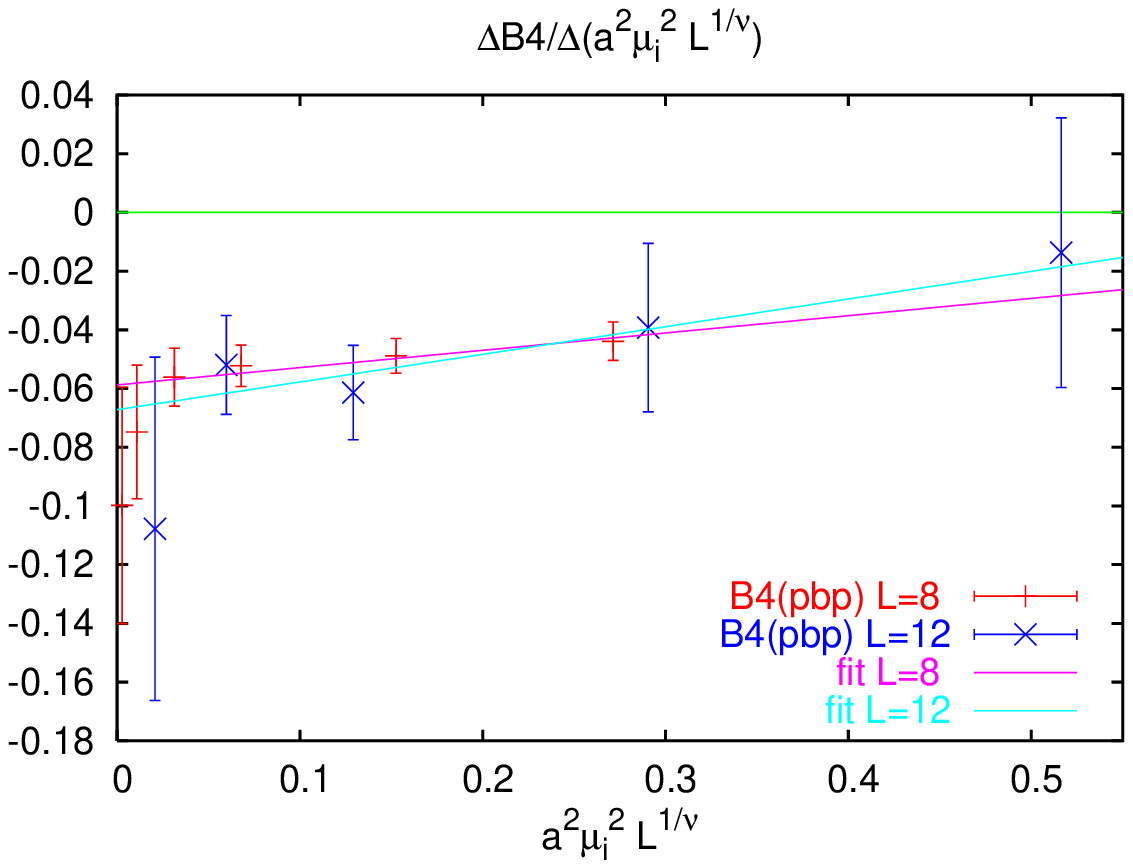}
\caption[]{\label{fss} Finite size scaling of the Taylor coefficients between $L=8, 12$ lattices.
For the pseudo-critical coupling, finite size corrections disappear in the thermodynamic
limit. $B_4$ is expected to scale according to \eq(\ref{scale}). 
}
\end{figure}

Next, we can check for finite volume scaling between our two volumes. This is done in \fig\ref{fss},
where the difference quotients are plotted against a volume-rescaled variable. The pseudo-critical
coupling has a finite thermodynamic limit, 
so its Taylor coefficients should become volume-independent
once the lattices are large enough. On the other hand,  
$B_4$ has to scale
in a way characteristic of the $3d$ Ising universality according to \eq(\ref{scale}). The rescaled curves in \fig\ref{fss} nearly
fall on top of each other, i.e.~both large volume scaling behaviours are realised to a good 
approximation between our $L=8$ and $L=12$ lattices.

\section{Simulations at imaginary $\mu$}
In this section we revisit our earlier attempt \cite{fp3} to determine the Taylor 
coefficients from simulations at non-vanishing imaginary $\mu=i\mu_i$, for which 
there is no sign problem. In this case the data contain no approximation 
other than the discretisation, and can be fitted by the truncated
series \eqs(\ref{bc},\ref{bseries}) directly. As in \cite{fp3},  
we consider an $8^3\times 4$ lattice, 
but we have added some more values of $\mu_i$ and increased the statistics.
We measured $B_4$ for 42 different pairs $(am,a\mu)$ drawn from 
$am\in[0.020,0.034], \mu_i\in[0,0.262]$,
each pair simulated at 3-5 different $\beta$-values around the pseudo-critical one.
Our overall statistics is well over 20 million trajectories.
 
Fits of \eqs(\ref{bc},\ref{bseries}) to various orders are collected in Table \ref{B4table}.
The first column specifies the fitting range, i.e.~the maximal imaginary chemical potential
included in the fits. We observe that $\beta_c$ can be fitted across the entire range, 
whereas for $B_4$ excellent fits are obtained up to $a\mu_i=0.245$. 
Once the fitting range is further increased, the $\chi^2$ abruptly grows, and the values
for the coefficients change. Clearly, this indicates the need for higher order terms.
Furthermore, for large $\mu_i$ there is a phase transition to an unphysical Z(3)-sector \cite{rw}, which
for $N_t=4$ happens at $a\mu_i=\frac{\pi}{3 N_t} \approx 0.262$. Hence, close to this point the scaling behaviour of this Z(3) transition
will mix into $B_4$, masking that of the transition we are after and leading to large finite volume effects.
For these reasons we discard fitting ranges $a\mu_i>0.245$.
This does not affect $\beta_c$, because it has a finite thermodynamic limit across the entire range.

\begin{table}[t]
\begin{center}
\hspace*{-1cm}
\begin{tabular}{|*{8}{r@{.}l|}l|}
\hline
\multicolumn{2}{|c|}{$|\mu|_{max}$} &
\multicolumn{2}{c|}{$\beta(am^c_0,0)$} &
\multicolumn{2}{c|}{$c_{10}$} &
\multicolumn{2}{c|}{$c_{20}$} &
\multicolumn{2}{c|}{$c_{01}$} &
\multicolumn{2}{c|}{$c_{02}$} &
\multicolumn{2}{c|}{$c_{11}$} &
\multicolumn{2}{c|}{$\chi^2/{\rm dof}$} \\
\hline
\hline
0&2    & 5&1354(1) &  1&95(3) & -14&1(2.7) & -0&75(1)  & 0&40(35) & -0&8(1.2) & 0&53 \\
0&2    & 5&1353(1) &  1&97(2) & -13&5(2.5) & -0&75(1)  & 0&43(35) & \none     & 0&52 \\
0&2    & 5&1352(1) &  1&90(2) & \none      & -0&763(6) & \none    & \none    & 1&14 \\
0&245  & 5&1354(1) &  1&96(3) & -14&4(2.1) & -0&735(7) & 0&84(12) & -0&31(78) & 0&52 \\
\hline
0&245  & 5&1354(1) &  1&97(2) & -14&5(2.1) & -0&736(7) & 0&84(11) & \none    & 0&51 \\
\hline
0&245  & 5&1348(2) &  1&89(2) & \none      & -0&783(5) & \none    & \none    & 2&17 \\
0&262  & 5&1354(1) &  1&96(3) & -15&2(2.3) & -0&729(8) & 1&00(11) & -0&61(86) & 0&64 \\
0&262  & 5&1354(1) &  1&98(2) & -15&4(2.3) & -0&729(7) & 0&99(11) & \none    & 0&63 \\
\hline
\hline
\multicolumn{2}{|c|}{$|\mu|_{max}$} &
\multicolumn{2}{c|}{$am^c$} &
\multicolumn{2}{c|}{$b_{10}$} &
\multicolumn{2}{c|}{$b_{20}$} &
\multicolumn{2}{c|}{$b_{01}$} &
\multicolumn{2}{c|}{$b_{02}$} &
\multicolumn{2}{c|}{$b_{11}$} &
\multicolumn{2}{c|}{$\chi^2/{\rm dof}$} \\
\hline
\hline
0&1    & 0&0259(4) & 14&0(7)   & \none         & 0&70(21) & \none      & \none         & 1&32\\
0&2    & 0&0257(4) & 14&0(1.5) & -404&8(119) & 0&38(58) & -7&2(14.2) & -63&4(54)   & 0&97\\
0&2    & 0&0259(4) & 14&0(7)   & \none         & 0&70(21) & \none      & \none        & 1&32\\
0&235  & 0&0257(4) & 14&7(1.3) & -256&5(99)  & 0&97(40) & 11&7(6.3)  & -6&9(28)    & 1&08\\
0&235  & 0&0257(4) & 14&9(6)   & -256&4(97)  & 0&98(39) & 11&9(6.1)  & \none        & 1&05\\
0&245  & 0&0257(4) & 14&6(1.3) & -255&7(93)  & 0&96(38) & 11&4(5.7)  & -8&1(27)    & 1&00\\
\hline
0&245  & 0&0257(4) & 14&9(6)   & -254&9(92)  & 0&96(37) & 11&6(5.6)  & \none        & 0&97\\
\hline
0&255  & 0&0255(4) & 15&6(1.5) & -321&1(118) & 1&46(46) & 22&2(6.8)  & 9&4(34)   & 1&59\\
0&255  & 0&0255(4) & 15&3(8)   & -322&4(116) & 1&47(46) & 22&1(6.6)  & \none        & 1&59\\
0&262  & 0&0254(5) & 16&5(1.6) & -311&4(126) & 1&96(47) & 32&2(6.7)  & -33&1(35)   & 1&98\\
0&262  & 0&0252(5) & 15&3(9)   & -308&8(126) & 2&02(48) & 32&4(6.7)  & \none        & 1&98\\
\hline 
\end{tabular}
\end{center}
\caption[]{Fits of the Taylor expansions $\beta_c(am,a\mu)$, \eq(\ref{bc}), (upper section) 
and $B_4(am,a\mu)$, \eq(\ref{bseries}), (lower section)
to data at imaginary $\mu$.
The first column specifies the fitting range.
}
\label{B4table}
\end{table}

For the good fits with $\chi^2/$dof $<1.5$, all coefficients
are stable within errors under variations of the fitting range.  Interestingly, we have a rather 
solid signal for a vanishing of the lowest order mixing between the $m$- and $\mu$-dependence in both
observables. We quote the highlighted fits as our results for the Taylor coefficients from fits
to imaginary $\mu$ data. 

The estimates for the leading terms agree within errors,  
while the subleading terms show some differences. The tendency for both coefficients is towards a smaller 
absolute value than that found by the derivative method. The extended
fitting ranges already suggest that this is due to the neglect of higher order terms.
Nevertheless, within one and a half standard deviations, also the sub-leading terms are consistent with those from the direct calculation of the derivatives, as is also visible in \fig\ref{der}.
We conclude that both methods considered here are able to provide the leading coefficient of
the Taylor series in $\mu$ quantitatively, as well as at least the sign of the sub-leading coefficients, though the derivative method is much more economical.

\section{Comparing and combining approaches}

We now have obtained compatible results from two independent methods
of calculating the leading Taylor coefficients of power series of critical parameters 
in $a\mu$. We stress again that
these are not merely two ways of analysing the same data. Rather, 
the simulations required for the two approaches also differ.  
Furthermore, the derivative method Sec.~\ref{delta} uses small chemical potentials
$|a \mu_i| \leq 0.1$, whereas the finite-$\mu_i$ simulations Sec.~4 consider 
larger  values $|a \mu_i|\geq 0.1$. The two approaches are thus complementary,
and it is reasonable to combine their results in order
to maximise the available information. To start with, the critical quark mass at zero density, 
$am^c_0$, and $b_{01}$ are best determined from the $\mu=0$ calculations of
Sec~\ref{delta}. Then, these values
can be used as input for the fits to a set of imaginary $\mu$ data. This is shown in 
Table \ref{B4table2}, where we have fixed $am^c_0=0.0259$ and $b_{01}=1.6$, according to our best estimates from $\mu=0$ simulations.

\begin{table}[t]
\begin{center}
\hspace*{-1cm}
\begin{tabular}{|*{6}{r@{.}l|}l|}
\hline
\multicolumn{2}{|c|}{$b_{10}$} &
\multicolumn{2}{c|}{$b_{20}$} &
\multicolumn{2}{c|}{$b_{02}$} &
\multicolumn{2}{c|}{$b_{03}$} &
\multicolumn{2}{c|}{$b_{21}$} &
\multicolumn{2}{c|}{$\chi^2/{\rm dof}$} \\
\hline
\hline
14&8(6)   & -421&7(177) & 49&4(8.3)  & 469&6(150) & -4950&5(4142)    & 1&05\\
\hline
14&9(6)   & -243&8(96)    & 47&4(8.2) & 421&9(145) & \none                         &  1&06\\
 \hline
 15&0(7)  & -208&3(184) & 23&5(1.4)  & \none                & -1470&7(4495)  & 1&33\\
 15&0(7)  & -156&8(101) & 23&7(1.3)  & \none                & \none                       & 1&31 \\
\hline
\end{tabular}
\end{center}
\caption[]{Fits of $B_4(am,a\mu)$, \eq(\ref{bseries}), to imaginary $\mu$ data, with the 
leading coefficients $am_0^c=0.0259, b_{01}=1.6$ fixed to their values determined at $\mu=0$.
}
\label{B4table2}
\end{table}

\begin{figure}[t]
\begin{center}
\includegraphics[width=0.6\textwidth]{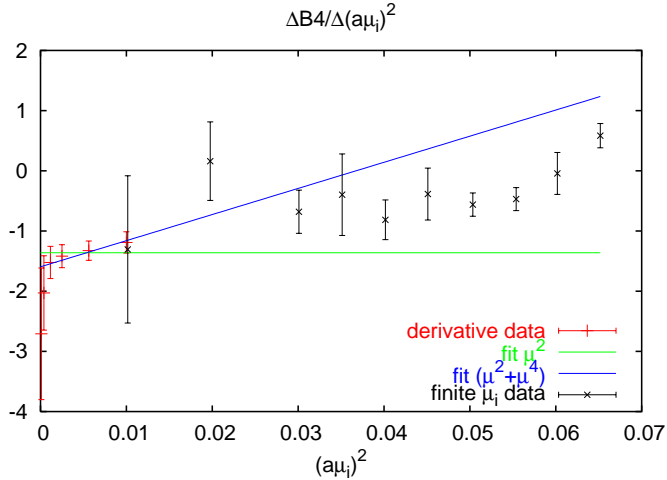}
\end{center}
\caption[]{\label{comb} Finite difference quotients on $L=8$ lattices for $am=0.0265$.
The left section $(a\mu_i)^2 \leq 0.01$ is determined via reweighting, 
the right section $(a\mu_i)^2 \geq 0.01$ via direct measurements of $B_4$
at finite $\mu=i\mu_i$.
The points at larger values clearly fall below the ${\cal O}(\mu^4)$ contribution, thus indicating a negative
$\mu^6$-term.
}
\end{figure}

With two coefficients fixed, we can actually consider the next order in $B_4$, i.e.~add 
terms ${\cal O}(m^3, m^2\mu^2,m\mu^4,\mu^6)$,  and still get sensible
fits. We find $b_{30}=b_{21}=0$ in all combinations of fit parameters. Removing the entirely unconstrained coefficients one by one, we observe that $b_{03}$ is
the most significant coefficient at this order. 
Note also that allowing for $b_{03}$, the value of $b_{02}$ relaxes
precisely to the value obtained from the derivative method, Table \ref{tabder}.
The same phenomenon is observed for the $\mu^2$ coefficient when releasing $\mu^4$.
Since $b_{01},b_{02},b_{03}$ all have the same sign, this is not surprising.
At imaginary $\mu=i\mu_i$ consecutive terms alternate in sign, and thus show a tendency
to cancel each other out. This demonstrates the 
"averaging" into lower order coefficients
of higher order terms that get significant but are excluded from a fit. 
For these reasons we believe the highlighted fit in Table \ref{B4table2}, with $b_{03}$ released,
to be the most trustworthy one.
Consistency between the two methods is also illustrated in \fig\ref{comb}, 
where we show the data for the reweighted
finite difference quotients together with those determined from the direct measurements of $B_4$
at finite $\mu=i\mu_i$.

We conclude that by combining both methods discussed here, we are sensitive to the $\mu^6$-term.
Its main effect is to give us some confidence in the order of magnitude of the $\mu^4$ term
determined in our earlier analyses. Moreover, without taking the value of $b_{03}$ too seriously,
it comes with the same sign as $b_{01}$ (see \fig\ref{comb}), thus leading to a negative $c_3'$. Hence, the
$\mu^6$-term also contributes to the shrinking of the first order region. 

\section{Conclusions}

For the continuum conversion, \eq(\ref{conv2}), we
need the change of temperature at the critical point with $\mu$. This is most conveniently extracted
by evaluating the pseudo-critical couplings at the critical points, 
fitting as a series in $(a\mu)^2$ and converting
to temperatures by the two-loop beta-function. 
Putting it all together, our final result reads\footnote{We note that the preliminary result 
$c_2=-12(6)$ reported earlier at various conferences \cite{prelim} corresponds to the value obtained
from fits to imaginary $\mu$ data alone.} 
\be
\frac{m_c(\mu)}{m_c(0)}=1-3.3(3)\left(\frac{\mu}{\pi T_c}\right)^2-47(20)\left(\frac{\mu}{\pi T_c}\right)^4
-\ldots
\ee
We conclude that our investigation of higher order contributions in 
$\mu^2$ fully supports and sharpens the conclusions of \cite{fp3}: 
the light mass range featuring a first order phase transition is 
shrinking with chemical potential. 
The quoted errors include one standard deviation uncertainty on the coefficients of the critical 
temperature, as well as on the parameters $m^c_0, b_{01}$ calculated at 
$\mu=0$.  
They do not include, however, the systematic error introduced by using the two-loop beta-function.

Nevertheless, our conclusion is stable against the use of another beta-function, as can be seen from 
\eq(\ref{conv2}).
The contribution of the second term to $c_1$ is always negative, irrespective of the beta-function used.
For $c_2$, the first term is numerically dominant and the second is guaranteed to be negative too,
no matter what the beta-function. The third term presently also gives a negative contribution, and it
is hard to see how a change in beta-function would change the sign and overhaul the contribution of the first two terms. Finally, we have good evidence that $c_3'$ is also
negative, i.e.~the $\mu^6$ term further shrinks the first order region, at least in lattice units.\footnote{We have not converted this coefficient to the continuum because
we cannot quantify all terms in \eq(\ref{conv2}) with sufficient accuracy at present.}  
Thus, all terms appear to have negative signs, confirming our conclusion irrespective of their precise
quantitative values. 
 
Our result therefore demonstrates beyond reasonable doubt 
that for $N_f=3$ the region of first order chiral phase transitions {\it shrinks} for 
small to moderate $\mu/T$, as in \fig\ref{schem} (right), and hence there is {\it no} 
chiral critical
point at chemical potentials $\mu \lsim T$ in a theory with three flavours of staggered fermions
and $m> m^c_0$
on an $N_t=4$ lattice. In order to draw conclusions for physical QCD, two further 
steps are necessary.
Firstly, the investigation has to be generalised to $N_f=2+1$. Secondly, and most importantly,
it has to be repeated on finer lattices,  since $N_t=4$ with $a\sim 0.3$ fm is too coarse to 
accurately reflect continuum physics. It is by no means excluded that cut-off effects are 
stronger than finite density effects on coarse lattices, and hence a change of sign in 
the curvature of the chiral critical surface in the continuum is a possibility. 
Simulations to address these issues are in progress. 
Finally, we emphasise that our findings only concern a {\em chiral} critical point, i.e. a point on the critical surface in \fig\ref{schem} bounding the region of first-order chiral transitions.
Our results do
not exclude an additional critical surface in \fig\ref{schem} (right), not analytically connected 
to the region of chiral phase transitions,
which could cause a critical point and phase transition of a different nature to exist.

\section*{Acknowledgements:}
This work is partially supported by the German BMBF, project
{\em Hot Nuclear Matter from Heavy Ion Collisions
and its Understanding from QCD}, No.~06MS254.
We thank the Minnesota Supercomputer Institute 
for providing computer resources,
and the CERN IT/GS group for their invaluable assistance and collaboration 
using the EGEE Grid for part of this project.
We acknowledge the usage of EGEE resources (EU project under contracts
EU031688 and EU222667). Computing resources have been contributed by a number
of collaborating computer centers, most notably HLRS Stuttgart (GER), NIKHEF (NL), CYFRONET (PL),
CSCS (CH) and CERN.
PdF thanks the Galileo Galilei Institute, Florence, and the Institute for
Nuclear Theory, University of Washington, Seattle, for the hospitality and
the stimulating atmosphere while part of this paper was written.

\end{document}